# High pressure effects on La(O,F)BiS$_2$ single crystal using diamond anvil cell with dual-probe diamond electrodes

Sayaka Yamamoto[1,2], Ryo Matsumoto[3]*, Shintaro Adachi[4], Yoshihiko Takano[1,2]

[1]*International Center for Materials Nanoarchitectonics (MANA),*

*National Institute for Materials Science, Tsukuba, Ibaraki 305-0047, Japan*

[2]*Graduate School of Pure and Applied Sciences, University of Tsukuba, 1-1-1 Tennodai, Tsukuba, Ibaraki 305-8577, Japan*

[3]*International Center for Young Scientists (ICYS),*

*National Institute for Materials Science, Tsukuba, Ibaraki 305-0047, Japan*

[4]*Nagamori Institute of Actuators, Kyoto University of Advanced Science, Ukyo-ku, Kyoto 615-8577, Japan*

E-mail: MATSUMOTO.Ryo@nims.go.jp

## Abstract

The high-pressure phase of La(O,F)BiS$_2$ exhibits the highest transition temperature among all the BiS$_2$-based superconductors. Various studies, such as investigation of isotope effects, have been conducted to explain its superconducting mechanism. However, there are very few reports on the electrical transport properties and vibration modes of single crystalline La(O,F)BiS$_2$ under high pressure. In this study, we developed a diamond anvil cell with dual-probe diamond electrodes to measure the electrical transport properties of La(O,F)BiS$_2$ single crystal and Pb as a manometer at low temperature. Using the developed system, a linear decrease in the transition temperature and phonon hardening was observed under the application of pressure on La(O,F)BiS$_2$ single crystal.

Following the discovery of BiS$_2$-based superconductors [1], various related compounds have been developed by changing the combination of conducting layers and charge reservoir blocking layers as $R$(O,F)Bi$Ch_2$ ($R$: La, Ce, Pr, Nd, Sm, Yb, $Ch$: S, Se) [2,3]. The properties of BiS$_2$-based materials are still being explored widely, as evidenced by the recent discovery of superconductivity in multi-layer type La$_2$O$_2$Bi$_3$Ag$_{0.6}$Sn$_{0.4}$S$_6$ with transition temperature ($T_c$) of 2.5 K [4,5]. The pairing mechanism of BiS$_2$-based superconductors is a subject of broad and current interest as a theoretical prediction of non-phonon mediated nature on LaO$_{0.5}$F$_{0.5}$BiS$_2$ [6] and experimental observation of an unconventional isotope effect in LaO$_{0.6}$F$_{0.4}$Bi(S,Se)$_2$ [7]. In addition to superconductivity, BiS$_2$-based materials have been studied as superior thermoelectric property with figure of merit $ZT$ = 0.36 on LaOBiSSe [8], platform of an exploration for superconductivity on high-entropy alloy [9], and candidate of topological superconductor [10].

High-pressure experiments play an important role in the early stages of development of BiS$_2$-based materials. Bi$_4$O$_4$S$_3$, the first BiS$_2$-based superconductor shows a gradual decrease in $T_c$ with application of high pressure [11]. On the other hand, the second one, La(O,F)BiS$_2$ exhibits a discrete enhancement in $T_c$ from 2.5 to 10.7 K with a structural phase transition from tetragonal to monoclinic [12]. Interestingly, the enhanced $T_c$ of the high-pressure phase can be quenched by high-pressure synthesis [13] or annealing [2] at 600−700°C and 2 GPa. Similar high-pressure effects with the structure change have been observed in various BiS$_2$-based superconductors, such as EuFBiS$_2$ [14] and Ce(O,F)BiS$_2$ [15], suggesting a common mechanism for $T_c$ enhancement. According to an investigation of isotope effects, the high-pressure phase of (Sr,La)FBiS$_2$ with a monoclinic crystal structure exhibits conventional-type $T_c$ shifts by substituting $^{32}$S and $^{34}$S [16], unlike the unconventional isotope effects observed in La(O,F)Bi(S,Se)$_2$ [7] and Bi$_4$O$_4$S$_3$ [17] with tetragonal structures. The high-pressure approach is also beneficial for inducing superconductivity in BiS$_2$-based compounds, such as NdO$_{0.8}$F$_{0.2}$Sb$_{1-x}$Bi$_x$Se$_2$ ($x$ = 0 to 0.8) [18] and EuFBiS$_2$ [19].

Study of high-pressure effects on the electrical transport properties of La(O,F)BiS$_2$ is especially important because it exhibits the highest $T_c$ among the BiS$_2$-based superconductors [3]. La(O,F)BiS$_2$ has a layered structure composed of conducting layers (BiS$_2$) and charge reservoir blocking layers (La(O,F)). Although the parent compound LaOBiS$_2$ is a semiconductor with direct band gap of 0.8-1.0 eV [20], electron doping by partial substitution of O by F induces superconductivity at 2.5 K under ambient pressure [2]. As mentioned before, $T_c$ can be increased by applying pressure in the case of polycrystalline samples [12]; however, there is no report on electrical transport measurements of single-crystal sample of La(O,F)BiS$_2$ under high pressure. In addition, the analysis of Raman vibration modes has never been conducted on La(O,F)BiS$_2$, which can provide important information about the structure under high pressure.

A diamond anvil cell (DAC) is typically used for simultaneous investigation of electrical transport properties and Raman modes under high pressure because a transparent anvil facilitates in situ optical analysis. However, precise determination of the pressure value at low temperatures is difficult in DAC. In general, the pressure in a DAC is estimated from the shift in the fluorescence peak position of a ruby crystal [21] in the sample chamber at room temperature. However, the pressure changes slightly at low temperatures owing to the thermal shrinkage of cell components. Moreover, detection of slight shifts in the fluorescence peak of ruby in the low-pressure region is difficult. In the case of the piston cylinder cell, the pressure at low temperatures can be estimated from the pressure-induced $T_c$ shift of Pb, which is located near the sample [22]. However, simultaneous measurements of the sample and Pb is difficult using the DAC because of the small size of the sample chamber.

In this study, we developed a DAC configuration with dual-probe electrodes composed of a boron-doped conducting diamond thin film [23,24] for simultaneous electrical transport measurements of two types of samples under high pressure. Using the developed system, electrical transport measurements and Raman spectroscopic studies of La(O,F)BiS$_2$ single crystal were carried out.

The diamond electrodes were designed by a nanofabrication technique using electron beam lithography and chemical vapor deposition, as described in the literature [25,26]. Figure 1 presents the schematic images of the DAC with dual-probe diamond electrodes. High pressure is generated by squeezing the diamond anvils. One side of the diamond anvil is equipped with boron-doped diamond electrodes for electrical transport measurements of the samples. An optical microscope image of the fabricated anvil is shown in Fig. 1 (c). The electrodes can be used repeatedly until the anvil is broken because the boron-doped diamond is mechanically and chemically stable and epitaxially grown from the anvil. The carrier concentration is of the order of $10^{21}$ cm$^{-3}$, indicating metallic nature with Fermi surface [23]. The other side of the diamond anvil has a culet with a diameter of 1 mm to induce high pressure. The sample chamber consists of a metal gasket made of stainless steel (316 L) with a hole of 600 μm diameter. As a pressure-transmitting medium, cubic boron nitride powder was added to the gasket hole. Raman spectrum of sample was acquired by irradiating the laser through the top diamond anvil, as shown in fig. 1(b). The electrodes were electrically insulated from the metal gasket using cubic boron nitride powders.

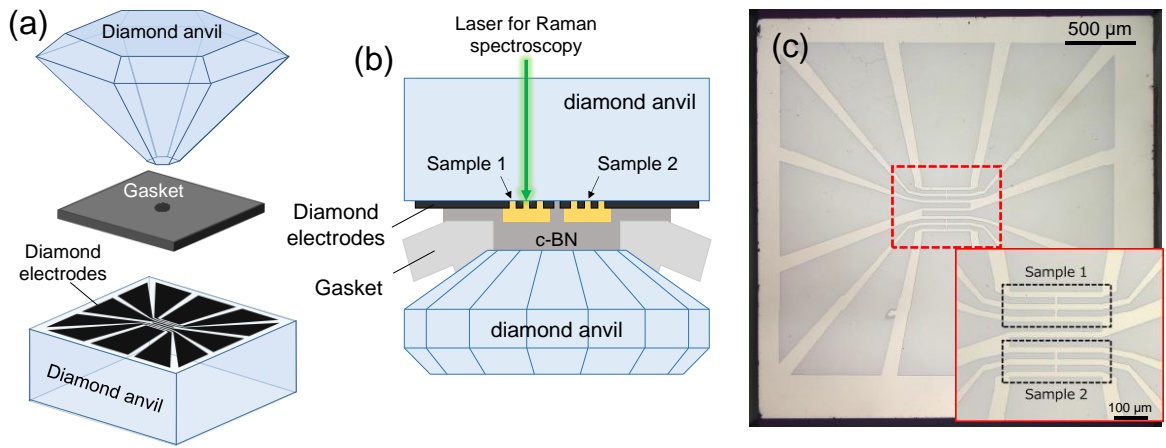

**Fig. 1.** Schematic image of the DAC with dual-probe diamond electrodes from (a) entire and (b) cross-sectional view. (c) Optical microscope image of the fabricated anvil.

Single crystals of La(O,F)BiS$_2$ were grown using an alkali metal flux with a nominal composition of LaO$_{0.5}$F$_{0.5}$BiS$_2$ according to a previously reported method [27]. The chemical compositions of the products were determined by scanning electron microscopy (SEM) equipped with energy dispersive X-ray spectroscopy (EDX) using a JSM-6010LA (JEOL) instrument, and single crystal X-ray diffraction using an XtaLAB mini (Rigaku) with Mo-K$\alpha$ radiation ($\alpha$ = 0.71072 Å). The temperature dependence of the resistance of La(O,F)BiS$_2$ in the range of 300 to 0.2 K at ambient pressure was measured by a four-probe method using an adiabatic demagnetization refrigerator (ADR) option on a physical properties measurement system (Quantum Design). The high-pressure generation and the in-situ transport measurements were performed using a DAC with dual-probe diamond electrodes. The Raman spectrum of the sample and the fluorescence spectrum of ruby were acquired using an inVia Raman microscope (RENISHAW).

Figure 2 (a) shows the typical SEM image of the obtained crystal exhibiting a well-developed plate-like shape. The compositional ratio was determined to be La:Bi:S = 1:0.99:1.72, by normalizing La = 1 in the EDX analysis, which is consistent with the cation composition of La(O,F)BiS$_2$. The single crystal XRD analysis revealed that the compound crystallizes with a tetragonal structure, with typical lattice constants of $a$ = 4.0541(15) Å and $c$ = 13.4826(52) Å. The EDX and XRD analyses establish that the obtained products are La(O,F)BiS$_2$ single crystals. The actual F concentration $x$ of the obtained LaO$_{1-x}$F$_x$BiS$_2$ is estimated to be $x$=0.2–0.3 from the known relationship between $x$ and the lattice constant of $c$ axis [28]. The crystal cleaved by a scotch tape was located on one side of the dual probe of the diamond anvil, as shown in Fig. 2 (b). Figure 2 (c) shows the enlarged image of the sample space. On the other side of the probe, 130 nm-thick Pb film was prepared by resistive heating vapor deposition. By measuring the $T_c$ of Pb, the applied pressure at low temperature was estimated using the relationship $P$ (GPa) = ($T_c^{ambient}$ – $T_c$) / 0.361 [22].

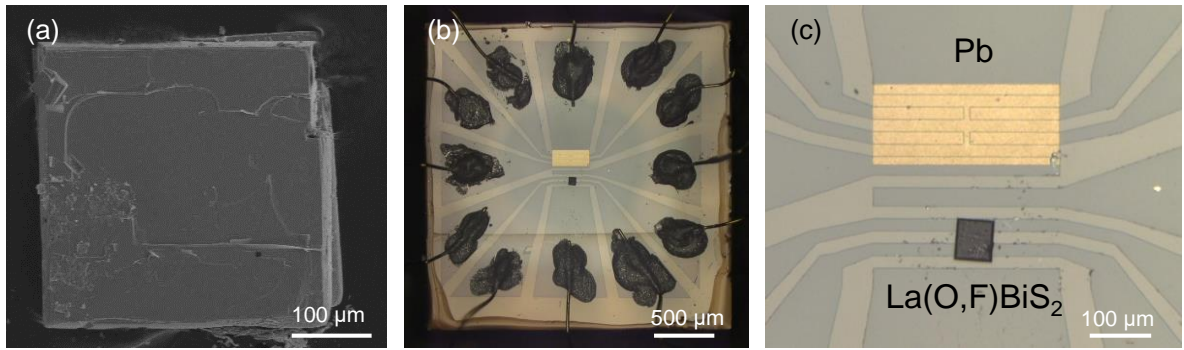

**Fig. 2.** (a) Typical SEM image of the obtained crystal. (b) Optical microscope image of the sample on the diamond anvil and (c) enlarged image at around the sample space.

Figure 3 (a) shows the temperature dependence of the normalized resistance of Pb under various pressures. The $T_c$ of Pb gradually decreases with increase in the pressure. The pressures estimated at low temperature from the $T_c$ of Pb are labeled in the figure. Here, the onset $T_c$ and zero-resistance $T_c$ of Pb mean the lowest pressure and highest pressure, respectively. The labeled pressures are average value of lowest and highest pressures. The superconducting transition becomes broader with increase of pressure, indicating an increase in pressure distribution. The temperature dependences of the normalized resistance of La(O,F)BiS$_2$ single crystals under different pressures are presented in Fig. 3 (b). The sample at ambient pressure exhibits an onset $T_c$ of 2.2 K. According to a previously reported relationship between $T_c$ and the amount of F ($x$) in the LaO$_{1-x}$F$_x$BiS$_2$ single crystal [28], the value of $x$ of the obtained crystal is estimated to be 0.2–0.3, which is consistent with the estimation from the lattice constant. By applying pressure, the onset $T_c$ is drastically enhanced to 8.9 K at 0.9 GPa. The transition width of superconductivity is broad because of the pressure distribution in the sample space. Above 0.9 GPa, the $T_c$ gradually decreases with increase in the pressure up to 2.2 GPa. In contrast, the zero-resistance $T_c$ is increased above 0.9 GPa, indicating an enhancement of volume fraction of higher $T_c$ phase.

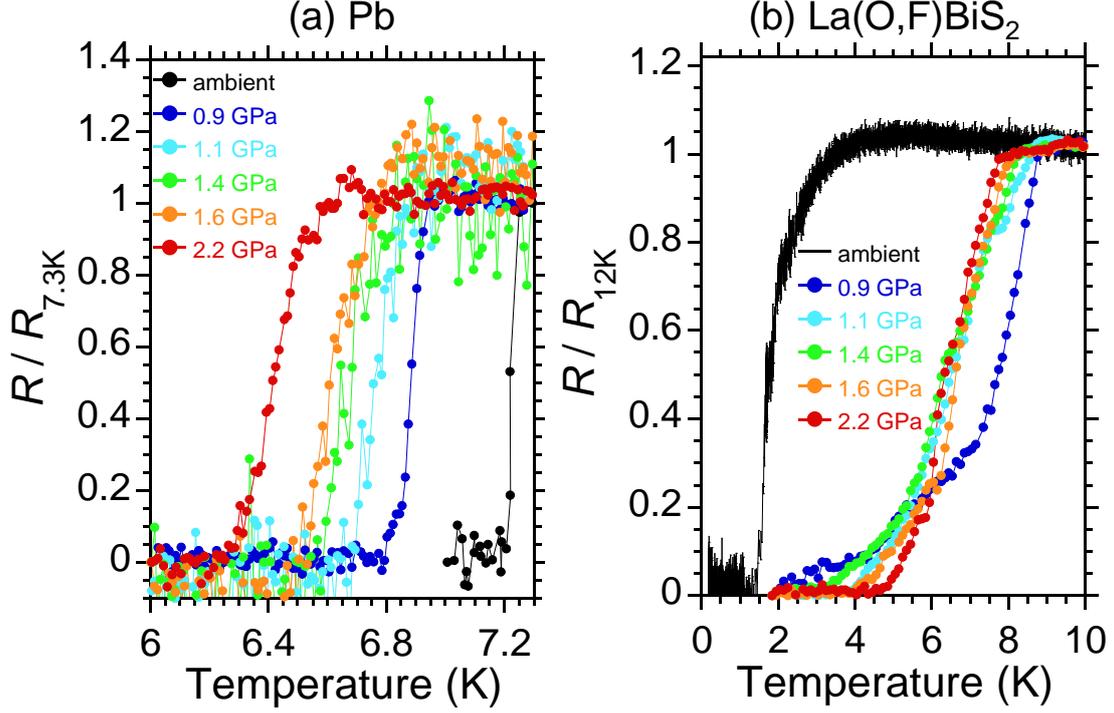

**Fig. 3.** Temperature dependence of the normalized resistance under various pressures on (a) Pb and (b) La(O,F)BiS$_2$ single crystal.

The circular plots in Fig. 4 (a) show the applied pressure dependence of $T_c$ of the La(O,F)BiS$_2$ single crystal. The rhombus symbols of $T_c^M$ and $T_c^{on}$ indicate the $T_c$ determined by magnetization and resistivity measurements, respectively, using a polycrystalline sample from a previous study [12]. A discrete enhancement in the $T_c$ is observed at 0.9 GPa. The $T_c$ values at pressures above 0.9 GPa of a single-crystal sample are slightly lower than those of a polycrystalline sample. According to the literature, La(O,F)BiS$_2$ synthesized with a lower content of F by high-pressure annealing exhibits a lower $T_c$ compared with that of an optimally doped sample [29]. Hence, the amount of F in our sample is believed to be lower than the optimum doping amount, which may result in a lower $T_c$ value. The discrete enhancement in $T_c$ at pressures above 0.9 GPa is followed by a linear suppression at the rate of $dT_c/dP = -0.71$ K/GPa up to 2.2 GPa, where $P$ is the applied pressure. Here we used the single crystal sample and hard pressure-transmitting medium of cubic boron nitride. This configuration generally induces a pseudo-uniaxial pressure [26]. An investigation of difference on distortion rate of lattice constant and bonding angle on single crystal under the pseudo-uniaxial pressure is impressive as a future work.

Figure 4 (b) shows the Raman spectra of the La(O,F)BiS$_2$ single crystal at various pressures. At ambient pressure, two peaks are observed at 70 and 123 cm$^{-1}$, which correspond to the Raman active $A_{1g}$ symmetric mode originating from the in-plane vibrations of Bi and S atoms [30]. According to the Raman investigation of Nd(O,F)BiS$_2$ [31], an electron-phonon coupling constant $\lambda \sim 0.16$ driven from the $A_{1g}$ mode is not sufficient to explain their $T_c$, suggesting an unconventional pairing mechanism at ambient pressure. The pattern of the vibration modes is completely changed at 0.9 GPa, indicating a pressure-induced structural phase transition. This is the first reported observation of the Raman modes of the higher $T_c$ phase of La(O,F)BiS$_2$ single crystal. Almost all the observed modes are gradually shifted to higher wavenumbers with increase in the pressure up to 2.2 GPa. The Raman spectrum of the sample showed the initial Raman modes after the pressure was released. Thus, it is evident that the pressure-induced phase transition and the phonon hardening are reversible. After the experiment, the diamond electrodes showed no degradation, and the same anvil was used repeatedly.

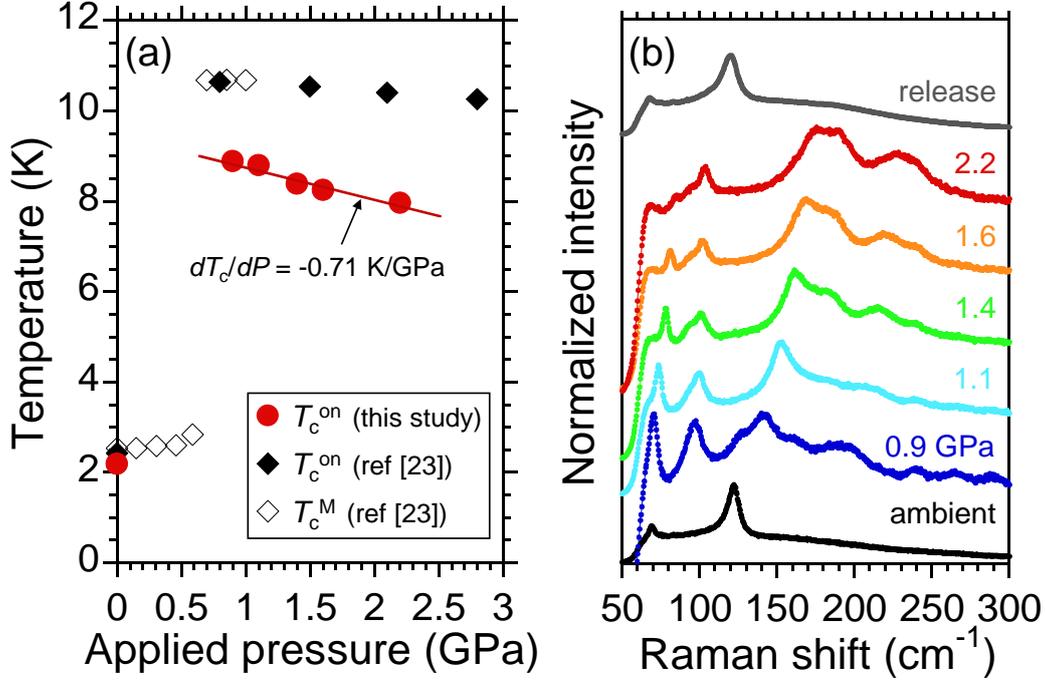

**Fig. 4.** (a) Applied pressure dependence of $T_c$ and (b) Raman spectra under various pressures on La(O,F)BiS$_2$ single crystal.

A linear decrease in $T_c$ with application of pressure is typically observed in conventional Bardeen-Cooper-Schrieffer (BCS)-type superconductors such as MgB$_2$ [32]. According to the BCS theory, the pressure effect on $T_c$ can be described by $d\ln T_c/dP = d\ln\omega/dP + 1.02/[\lambda(1-\mu^*)-\mu^*]^2(d\lambda/dP)$, where $P$ is the applied pressure, $\omega$ is the phonon frequency, $\mu^*$ is the Coulomb repulsion, and $\lambda$ is the electron-phonon coupling constant [32,33]. Here, $\lambda$ is $N(0)\times\langle I^2\rangle/M\langle\omega^2\rangle$, where $N(0)$ is the density of states at the Fermi energy, $\langle I^2\rangle$ is the averaged square of the electronic matrix element, $M$ is the atomic mass, and $\langle\omega^2\rangle$ is the averaged square of the phonon frequency. The applied pressure is expected to reduce $N(0)$ and enhance $\omega$ because of bandwidth broadening and phonon hardening, respectively. Although the enhancement of $\omega$ gives a positive effect for the first term $d\ln\omega/dP$, the decrease in $N(0)$ and increase in $\langle\omega^2\rangle$ induce a drastic decrease in $\lambda$, namely a negative effect for the second term $d\lambda/dP$, possibly resulting in $T_c$ reduction. The linear decrease in $T_c$ and the phonon hardening feature from the Raman spectra of the high-pressure phase of La(O,F)BiS$_2$ support phonon-mediated pairing mechanism of superconductivity, as observed in the conventional isotope effect [16].

In conclusion, we successfully demonstrated simultaneous measurements of the electrical transport properties of La(O,F)BiS$_2$ single crystal and Pb as a manometer at low temperature by using the developed DAC with dual-probe diamond electrodes. The transport measurements and the Raman studies under high pressures revealed that the discrete enhancement in $T_c$ originated from the structural phase transition. A linear reduction in $T_c$ and phonon hardening were observed with increase in the applied pressure on the high-pressure phase of the La(O,F)BiS$_2$ single crystal. The developed DAC with dual-probe electrodes is extremely useful for high-pressure research because two samples can be measured simultaneously with in situ optical analysis.

**Acknowledgments**
This work was partly supported by JST CREST Grant No. JPMJCR16Q6, JST-Mirai Program Grant Number JPMJMI17A2, JSPS KAKENHI Grant Number JP19H02177, 20H05644, and 20K22420. A fabrication of diamond was supported by NIMS Nanofabrication Platform in Nanotechnology Platform Project sponsored by the Ministry of Education, Culture, Sports, Science and Technology (MEXT). The authors would like to acknowledge the ICYS Research Fellowship, NIMS.